\documentclass[prd,twocolumn,showpacs,preprintnumbers,amsmath,nofootinbib,amssymb,floatfix]{revtex4-1}
\usepackage{graphicx}
\usepackage{epsfig,float}
\usepackage[sort&compress]{natbib}
\usepackage{subfigure}
\usepackage{amsmath}
\usepackage{amsfonts}
\usepackage{cancel}
\usepackage{amssymb}
\usepackage{hyperref}
\usepackage{color}
\begin{document}
\arraycolsep1.5pt

\title{Looking for a hidden-charm pentaquark state with strangeness $S=-1$ from $\Xi^-_b$ decay into $J/\psi K^- \Lambda$}

\author{Hua-Xing Chen}
\affiliation{School of Physics and Nuclear Energy Engineering and
International Research Center for Nuclei and Particles in the
Cosmos, Beihang University, Beijing 100191, China}

\author{Li-Sheng Geng} \email{lisheng.geng@buaa.edu.cn}
\affiliation{School of Physics and Nuclear Energy Engineering and
International Research Center for Nuclei and Particles in the
Cosmos, Beihang University, Beijing 100191, China}
\affiliation{State Key Laboratory of Theoretical Physics, Institute
of Theoretical Physics, Chinese Academy of Sciences, Beijing 100190,
China}

\author{Wei-Hong Liang}
\affiliation{Department of Physics, Guangxi Normal University,
Guilin 541004, China}

\author{Eulogio Oset}
\affiliation{Institute of Modern Physics, Chinese Academy of
Sciences, Lanzhou 730000, China}\affiliation{Departamento de
F\'{\i}sica Te\'orica and IFIC, Centro Mixto Universidad de
Valencia-CSIC Institutos de Investigaci\'on de Paterna, Aptdo.
22085, 46071 Valencia, Spain}

\author{En Wang}
\affiliation{Department of Physics, Zhengzhou University, Zhengzhou, Henan 450001, China} 

\author{Ju-Jun Xie}
\affiliation{Institute of Modern Physics, Chinese Academy of
Sciences, Lanzhou 730000, China} \affiliation{Research Center for
Hadron and CSR Physics, Institute of Modern Physics of CAS and
Lanzhou University, Lanzhou 730000, China} \affiliation{State Key
Laboratory of Theoretical Physics, Institute of Theoretical Physics,
Chinese Academy of Sciences, Beijing 100190, China}
\date{\today}

\begin{abstract}

Assuming that the recently observed hidden-charm pentaquark state, $P_c(4450)$, is of molecular nature as predicted
in the unitary approach, we propose to study the  decay of $\Xi^-_b\rightarrow J/\psi K^- \Lambda$ to search for
the strangeness counterpart of the $P_c(4450)$. There are three ingredients in the decay mechanism:
the weak decay mechanism, the hadronization mechanism, and the finite state interactions in the meson-baryon system of 
strangeness $S=-2$ and isospin $I=1/2$ and of the $J/\psi\Lambda$. All these have been tested extensively.
 As a result, we provide a genuine prediction of the differential cross section where 
a strangeness hidden-charm pentaquark state, the counterpart of the $P_c(4450)$, can be clearly seen.  The decay rate is estimated to
be of similar magnitude as the $\Lambda_b^0\rightarrow J/\psi K^- p$ observed by the LHCb collaboration.
\end{abstract}

\maketitle
\section{Introduction}
Recently, the LHCb collaboration has reported on the observation of two hidden-charm pentaquark  states in
$\Lambda_b^0\rightarrow J/\psi K^- p$ decays,  
the $P_c(4380)$ with a mass of $4380\pm8\pm29$ MeV and a width of $205\pm18\pm86$
MeV and the $P_c(4450)$ with a mass of 
$4449.8\pm1.7\pm2.5$ MeV and a width of $39\pm5\pm19$ MeV~\cite{Aaij:2015tga}.  The preferred $J^P$ assignments are of opposite parity. The best
fit  yields  spin-parity $J^P$ values of $(3/2^-,5/2^+)$, but other possibilities,
either $(3/2^+,5/2^-)$ or $(5/2^+,3/2^-)$, are also acceptable.  The decay branching ratios  $\Lambda^0_b\rightarrow P_c^+ K^-$ and $P_c^+\rightarrow J/\psi p$ have
been measured as well~\cite{Aaij:2015fea}. In the past decade, many
mesonic exotic states have been observed experimentally and many of them clearly contain more than the minimum quark content dictated by the naive quark model, 
such as the charged $Z_c(4430)$~\cite{Choi:2007wga,Chilikin:2013tch,Aaij:2014jqa} 
 and $Z_c(3900)$~\cite{ Ablikim:2013mio,Liu:2013dau,Xiao:2013iha,BESIII:2015kha} states. The $P_c$ states
 are the first exotic states observed in the heavy-flavor baryonic sector.

The observation of the $P_c$ states has aroused a lot of interest in the theoretical community. They have been
studied in various frameworks, such as  the  molecular picture~\cite{ Chen:2015loa,Roca:2015dva,He:2015cea,Meissner:2015mza},
the diquark picture~\cite{Lebed:2015tna,Maiani:2015vwa,Anisovich:2015cia,Ghosh:2015ksa,latest:diquark}, the QCD sum rules~\cite{Chen:2015moa,Wang:2015epa},
and  the soliton model~\cite{Scoccola:2015nia}.
On the other hand, questions have been raised regarding whether the observed enhancement could be due to kinematical effects or singularities~\cite{Guo:2015umn,Liu:2015fea,Mikhasenko:2015vca}.
 In a recent publication, it was suggested that the existence of exotic
 $cs\bar{c}\bar{u}$ states can imitate broad bumps in the $J/\psi p$ invariant mass distributions and thus affects the interpretation of the $P_c(4380)$~\cite{Anisovich:2015xja}. 
 Further discussions on the nature of the two $P_c$ states  can be found in Refs.~\cite{Mironov:2015ica,Burns:2015dwa}.
 
 One should note that even before the LHCb announcement, the existence of hidden-charm pentaquark states have been explored
 both in the molecular picture~\cite{Wu:2010jy,Wu:2010vk,Yang:2011wz,Xiao:2013yca, Karliner:2015ina} and in the
quark models~\cite{Wang:2011rga,Yuan:2012wz}).~\footnote{For a list of early references on pentaquark states,
see, e.g., Ref.~\cite{Stone:2015iba}.} The discovery potential of such states has been explored in $\gamma$~\cite{Huang:2013mua}
and $\pi$~\cite{Garzon:2015zva} induced reactions.

It is clear that at this moment, both the experimental and theoretical studies
are not yet conclusive. Experimentally,  the ambiguities in the spin-parity assignments should be
clarified. Theoretically, different interpretations are often not consistent with each other, not only in terms of
the dominant Fock components of these states but also in terms of the predicted or obtained spin-parities. The only way to clarify the
situation, given the rather large statistics already achieved by LHCb, is to study complementary reactions or decay modes where the $P_c$ states or
their counterparts (predicted by various models) can be observed.

In Refs.~\cite{ Wang:2015jsa,Kubarovsky:2015aaa,Karliner:2015voa}, photoproduction of the $P_c$ states off a proton target have been studied.  In Ref.~\cite{Cheng:2015cca},
assuming the $P_c$ states are genuine pentaquark states belonging to either an octet or a decuplet representation, the decays of $\Lambda_b^0$, $\Xi^0_b$, and $\Xi^-_b$ into a pentaquark state and
a pseudoscalar meson have been studied.  The  decays of $b$-baryons into a pentaquark state and a pseudoscalar meson have also been examined in the diquark model~\cite{Li:2015gta}.
Experimental studies of either the photoproductions or the decay modes of $b$-baryons into all available final states will definitely help us better understand the nature of the pentaquark states.

One should note that most theoretical approaches
predicted the existence of the $P_c$ counterparts. In particular, in the unitary approach of Ref.~\cite{Wu:2010vk} that in addition to an isospin 1/2 and strangeness zero state, 
two more states are predicted in the isospin zero and strangeness $-1$ sector. If the $P_c(4450)$ state corresponds to the
non strange state(s) as shown in Ref.~\cite{Roca:2015dva},  there is good reason to believe that its strange counterparts exist as well. The hidden-charm pentaquark states
with strangeness can decay into $J/\psi\Lambda$. Therefore, in the present work, we propose to study the decay of the $\Xi_b^-$ state into $J/\psi K^- \Lambda$.  The mechanism of this reaction resembles that of the $\Lambda_b^0\rightarrow
J/\psi K^- p$ suggested in Ref.~\cite{Roca:2015dva}, which can naturally explain the LHCb data using the input from the
unitary model of Refs.~\cite{Xiao:2013yca,Wu:2010vk}. The 
experimental observation or invalidation of the existence of such a state can help clarify the nature of the $P_c$ states and improve our understanding of low-energy strong interactions.

\section{Formalism}\label{sec:formalism}
In this section, we describe the  weak decay process of $\Xi^-_b\to J/\psi \Lambda K^-$. Following the  formalism first proposed in Ref.~\cite{Liang:2014tia} and also used to
study the $\Lambda_b^0\rightarrow J/\psi K^- p$ decay~\cite{Roca:2015dva,Roca:2015tea}, we can separate  the decay of the $\Xi^-_b$ into two steps, weak decay and hadronization, and final state interactions. 
\subsection{Weak decay and  hadronization}
\begin{figure}[!htb]
\centering
  \includegraphics[width=0.45\textwidth]{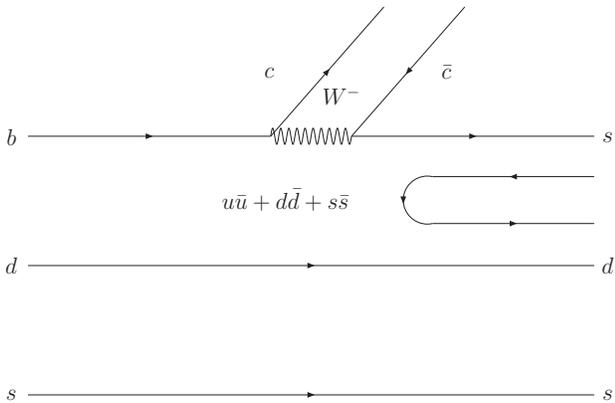}
\caption{Weak decay and hadronization mechanism of $\Xi^-_b\to J/\psi \Lambda K^-$. }
  \label{fig:mechanism}
\end{figure}

At the quark level,
the decay of $\Xi^-_b\rightarrow J/\psi K^- p$ is depicted in Fig.~\ref{fig:mechanism}. The quark content of $\Xi^-_b$ is $bds$, and the $d$ and $s$ quarks 
are in a state of spin zero. To have the color degree antisymmetric, the flavor part of the wave function should be antisymmetric with respect to $d$ and $s$. As a result, the $\Xi^-_b$ wave function 
can be written as
\begin{equation}
\Xi^-_b=|b\rangle|ds\rangle\Rightarrow\frac{1}{2}|b\rangle(|d\rangle|s\rangle-|s\rangle|d\rangle)=\frac{1}{2}b(ds-sd).
\end{equation}
In the second equality, we have adopted a simplified notation for the wave functions.
In Fig.~\ref{fig:mechanism}, the $b$ quark first decays into a $c$ quark by emitting a $W^-$ meson, then the $W^-$ translates into a pair of $\bar{c}$ and $s$ quarks,
which is Cabibbo favored.  The pair of $c\bar{c}$ hadronizes into
the $J/\psi$, while the $s$ quark picks up an anti-quark from the vacuum to form a $\bar{K}$ meson or an $\eta$ meson, the spectator pair $ds$ then hadronizes into a baryon with the remaining quark from
the vacuum. In the following, we have to find out how the quark combinations $Q=s(\bar{u}u+\bar{d}d+\bar{s}s)\frac{1}{2}(ds-sd)$ hadronizes into a pair of ground state meson and baryon.

 The hadronization into a meson baryon pair can be
achieved by replacing the $M=q\bar{q}$ matrix in SU(3) flavor space with its counterpart $\phi$ using hadronic degrees of freedom, namely
\begin{equation}
M=\left(\begin{array}{ccc}u\bar{u}&u\bar{d}&u\bar{s}\\
d\bar{u} & d\bar{d} & d\bar{s} \\
s\bar{u} & s\bar{d} & s\bar{s}\end{array}\right)\rightarrow
\phi=\left(\begin{array}{ccc}\frac{\pi^0}{\sqrt{2}}+\frac{\eta}{\sqrt{3}} & \pi^+ & K^+\\
\pi^- & -\frac{\pi^0}{\sqrt{2}}+\frac{\eta}{\sqrt{3}} & K^0\\
K^- & \bar{K}^0 & -\frac{\eta}{\sqrt{3}}\end{array}\right),
\end{equation}
where we have used standard $\eta/\eta'$ mixing~\cite{Bramon:1992kr} and have neglected the $\eta'$ because of its heavy mass.  With such a replacement, we obtain
\begin{eqnarray}
Q&=&\sum\limits_{i=1}^3 M_{3i}q_i\frac{1}{\sqrt{2}}(ds-sd)=\sum\limits_{i=1}^3\phi_{3i}q_i\frac{1}{\sqrt{2}}(ds-sd)\nonumber\\
&=& K^-\left[ \frac{1}{\sqrt{2}}u(ds-sd)\right]+\bar{K}^0 \left[ \frac{1}{\sqrt{2}}d(ds-sd)\right]\nonumber\\
&&-\frac{\eta}{\sqrt{3}}\left[ \frac{1}{\sqrt{2}}s(ds-sd)\right].
\end{eqnarray}
The combinations of three quarks can be written in terms of the ground-state baryon wave functions with a bit of algebra~\cite{Close:1979bt}
\begin{equation}\label{eq:hadronization}
Q=\sqrt{\frac{3}{2}}\bar{K}\Sigma|_{I=1/2}-\sqrt{\frac{1}{6}}\bar{K}\Lambda -\sqrt{\frac{1}{3}}\eta\Xi,
\end{equation}
where we have introduced the isospin 1/2 combination $\bar{K}\Sigma|_{I=1/2}=\sqrt{\frac{2}{3}}(\frac{1}{\sqrt{2}}K^-\Sigma^0+\bar{K}^0\Sigma^-)$, $\bar{K}\Lambda=K^- \Lambda$, and $\eta\Xi=\eta\Xi^-$. The process
described above corresponds to the tree-level Feynman diagram of  Fig.~\ref{fig:decaymodel}(a).

Other hadronization processes, for instance, that of  the $d$ or $s$ quark of the $\Xi^-_b$ hardronizing into a meson is penalized compared to the one described above,
 because of the involvement of large momentum transfer~\cite{Miyahara:2015cja}. 

\subsection{Final state interactions }
\begin{figure}[!htb]
\centering
  \includegraphics[width=0.45\textwidth]{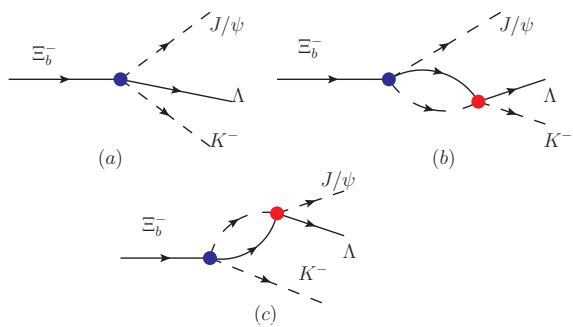}
\caption{(Color online) Feynman diagrams for $\Xi^-_b\to J/\psi \Lambda K^-$ decay: (a) direct $J/\psi \Lambda K^-$ vertex at tree level, (b) final state interaction of $\bar K\Lambda $, and (c) final state interaction of $J/\psi \Lambda$.} 
  \label{fig:decaymodel}
\end{figure}

In addition to the tree level diagram [Fig.~\ref{fig:decaymodel}(a)], we need to take into account the final state interactions of these meson-baryon pairs, which are known to be very
strong and are depicted in Fig.~\ref{fig:decaymodel}(b) and (c). The amplitude $\mathcal{M}(M_{J/\psi\Lambda},M_{\bar K\Lambda })$ for the transition can be written as, 
\begin{eqnarray}
&&\mathcal{M}(M_{J/\psi\Lambda}, M_{\bar K\Lambda })\nonumber \\&=& V_p\left[ h_{ \bar K\Lambda}+\sum_{i}h_iG_i(M_{\bar  K\Lambda})\,t_{i,\bar K\Lambda }(M_{ \bar K\Lambda}) \right. \, \nonumber \\
&& \left. \phantom{\sum_{i}}\hspace*{-0.2cm}  +h_{\bar K\Lambda }G_{J/\psi\Lambda}(M_{J/\psi\Lambda})\,t_{J/\psi\Lambda,J/\psi\Lambda}(M_{J/\psi\Lambda})\right] \, \nonumber \\
&=&V_p\left( h_{\bar  K\Lambda} + T_{\bar  K\Lambda} + T_{J/\psi\Lambda}\right) ,\label{eqn:fullamplitude}
\end{eqnarray}
where $V_p$ expresses the hadronization strength, and
 $G_i$  (the channel indices $i=\bar K \Lambda, ~\bar K \Sigma,~\eta\Xi$) denotes the one-meson-one-baryon loop function, chosen in accordance with the unitary model for the scattering matrix $t_{ij}$ 
 that will be described in the following subsection. $M_{\bar{K}\Lambda}$ and  $M_{J/\psi\Lambda}$ are the invariant masses of the final states $K^-\Lambda $ and $J/\psi\Lambda$, respectively, and $h_i$ stands for the weights of the transition, which are given by Eq.~(\ref{eq:hadronization}),
\begin{gather}
h_{\bar K\Lambda}= -\sqrt{\frac{1}{6}},~~ 
h_{\bar K \Sigma}=\sqrt{\frac{3}{2}},~~
h_{\eta\Xi}=-\sqrt{\frac{1}{3}}. 
\end{gather}

Final state interactions between a ground state octet baryon and a pseudoscalar meson in the strangeness $-2$ and isospin $1/2$ channel
has been studied in detail in the unitary model of Ref.~\cite{Ramos:2002xh}.  In this work, a pole is found on the complex plane and identified as
the experimentally observed $\Xi(1620)$. A subsequent work along the same lines showed that in addition to the $\Xi(1620)$ state also the $\Xi(1690)$ was
generated~\cite{Gamermann:2011mq}. In the present work, we choose this approach to describe the interactions among the coupled channels $\bar{K}\Lambda$, $\bar{K}\Sigma$, and $\eta\Xi$.

\subsection{Chiral unitary model for the meson baryon interaction in $S=-2$}

In the unitary approach of Ref.~\cite{Ramos:2002xh}, the transition amplitudes are written as
\begin{equation}
t=\left[1-VG\right]^{-1}V, \label{eq:tvg}
\end{equation} 
where the matrix $V$ is obtained from the lowest order meson baryon chiral Lagrangian,
\begin{eqnarray}
V_{ij}(I=1/2)&=&-C_{ij}\frac{1}{4f^2}(2\sqrt{s}-M_i-M_j)\nonumber \\
&&\times \left(\frac{M_i+E_i}{2M_i}\right)^{1/2} \left(\frac{M_j+E_j}{2M_j}\right)^{1/2},
\end{eqnarray}
where the magnitudes $E_i$ and $M_i$ are the energy and mass of the baryon in channel $i$,  
and the coefficients $C_{ij}$ are shown in Table~\ref{tab:cij}. The optimal choice for the decay constant $f=1.123f_\pi$ is used, and $f_\pi=93$ MeV~\cite{Ramos:2002xh}. 
\begin{table}[ht]
\centering \caption{\small Coefficients $C_{ij}$ of the meson
baryon amplitudes for isospin $I=1/2$ ($C_{ji}=C_{ij}$)~\cite{Ramos:2002xh}.} \vspace{0.5cm}
\begin{tabular}{l|rrrr}\hline \hline
         &  $\pi\Xi$ & ${\bar K} \Lambda$ &
${\bar K} \Sigma$ & $\eta\Xi$ \\
        \hline
        $\pi\Xi$ & 2 & $-3/2$ & $-1/2$ & 0\\
 ${\bar K}\Lambda$    & &
0 & 0 &
$-3/2$  \\

${\bar K}\Sigma$ & &  &
2 &
$3/2$ \\
 
$\eta\Xi$   & &  &  & 0 \\ 
\hline \hline
\end{tabular}
\label{tab:cij}
\end{table}

The (diagonal) matrix $G$ in Eq.~(\ref{eqn:fullamplitude}) and Eq.~(\ref{eq:tvg}) accounts for the loop
integral of a meson and a baryon propagator and 
depends on the regularization scale, $\mu$, and a subtraction constant for
each channel, $a_l$, that comes
from a subtracted dispersion relation. The
analytical expression of $G$ is given as follows~\cite{Oset:2001cn},
\begin{eqnarray} 
G_{l} &=& i 2 M_l \int \frac{d^4 q}{(2 \pi)^4} \,
\frac{1}{(P-q)^2 - M_l^2 + i \epsilon} \, \frac{1}{q^2 - m^2_l + i
\epsilon}  \nonumber \\ &=& \frac{2 M_l}{16 \pi^2} \left\{ a_l(\mu) + \ln
\frac{M_l^2}{\mu^2} + \frac{m_l^2-M_l^2 + s}{2s} \ln \frac{m_l^2}{M_l^2} +
\right. \nonumber \\ & &  \phantom{\frac{2 M}{16 \pi^2}} +
\frac{\bar{q}_l}{\sqrt{s}} 
\left[ 
\ln(s-(M_l^2-m_l^2)+2\bar{q}_l\sqrt{s})\right. \nonumber \\
&& \left. \phantom{\frac{2 M}{16 \pi^2} +
\frac{\bar{q}_l}{\sqrt{s}}} 
 \hspace*{-0.3cm}
+\ln(s+(M_l^2-m_l^2)+2\bar{q}_l\sqrt{s}) \right. \nonumber  \\
& & \left. \phantom{\frac{2 M}{16 \pi^2} +
\frac{\bar{q}_l}{\sqrt{s}}} 
 \hspace*{-0.3cm}- \ln(-s+(M_l^2-m_l^2)+2\bar{q}_l\sqrt{s})\right. \nonumber \\
&&\left. \phantom{\frac{2 M}{16 \pi^2} +
\frac{\bar{q}_l}{\sqrt{s}}} 
\left. \hspace*{-0.3cm} -\ln(-s-(M_l^2-m_l^2)+2\bar{q}_l\sqrt{s}) \right]
\right\} \ ,
\label{eq:gpropdr}
\end{eqnarray}        
where $M_l$ and $m_l$ are the masses of baryon and meson in the $l$-th channel, and the regularization scale $\mu=630$ MeV~\cite{Oset:2001cn} and the subtraction constant $a_l=-2.0$\footnote{We have checked that Set~5 ($a_{\pi\Xi}=-3.1$, $a_{\bar{K}\Lambda}=-1$, $a_{\bar{K}\Sigma}=-2$ and $a_{\eta\Xi}=-2$) in Table~2 of Ref.~\cite{Ramos:2002xh} gives similar results as our choice.} are used for $\pi\Xi$, $\bar{K}\Lambda$, $\bar{K}\Sigma$, 
and $\eta\Xi$ channels. For the $J/\psi\Lambda$ channel, we take $\mu=1000$ MeV and $a_l=-2.3$~\cite{Wu:2010vk}.

\begin{figure}[!htb]
\centering
  \includegraphics[width=0.45\textwidth]{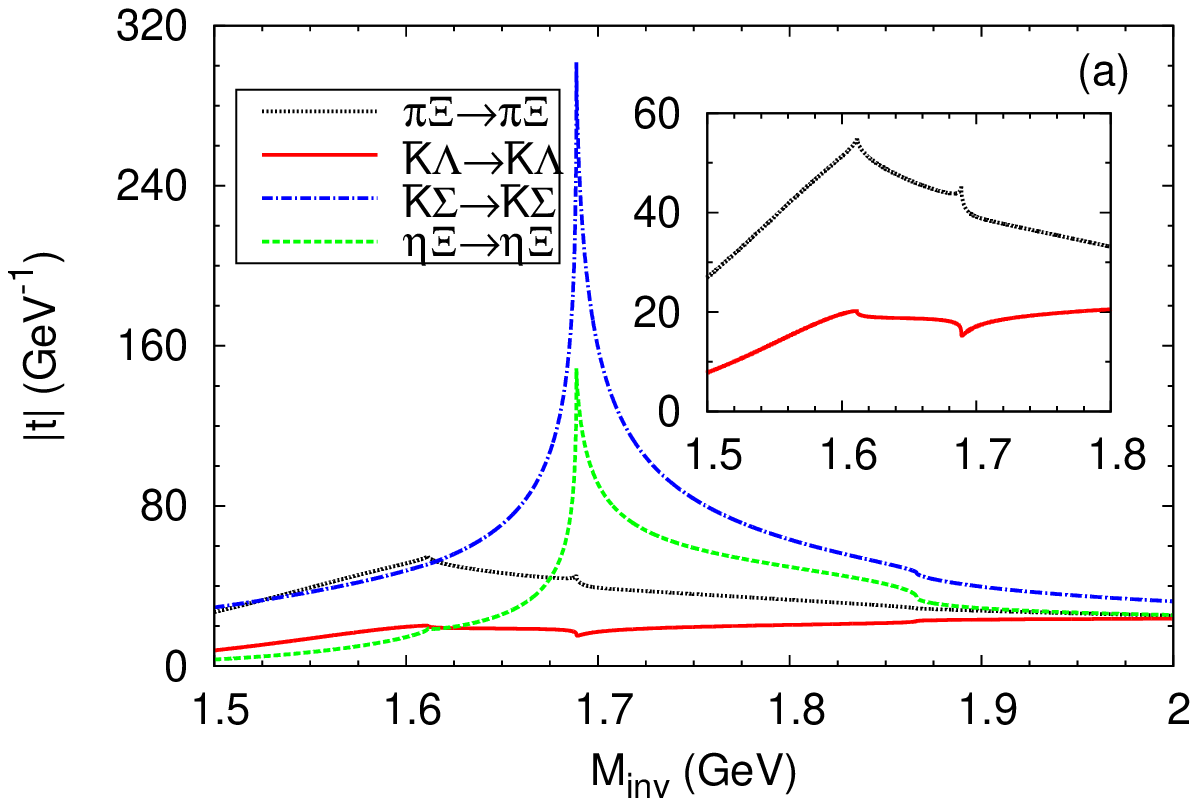}
    \includegraphics[width=0.45\textwidth]{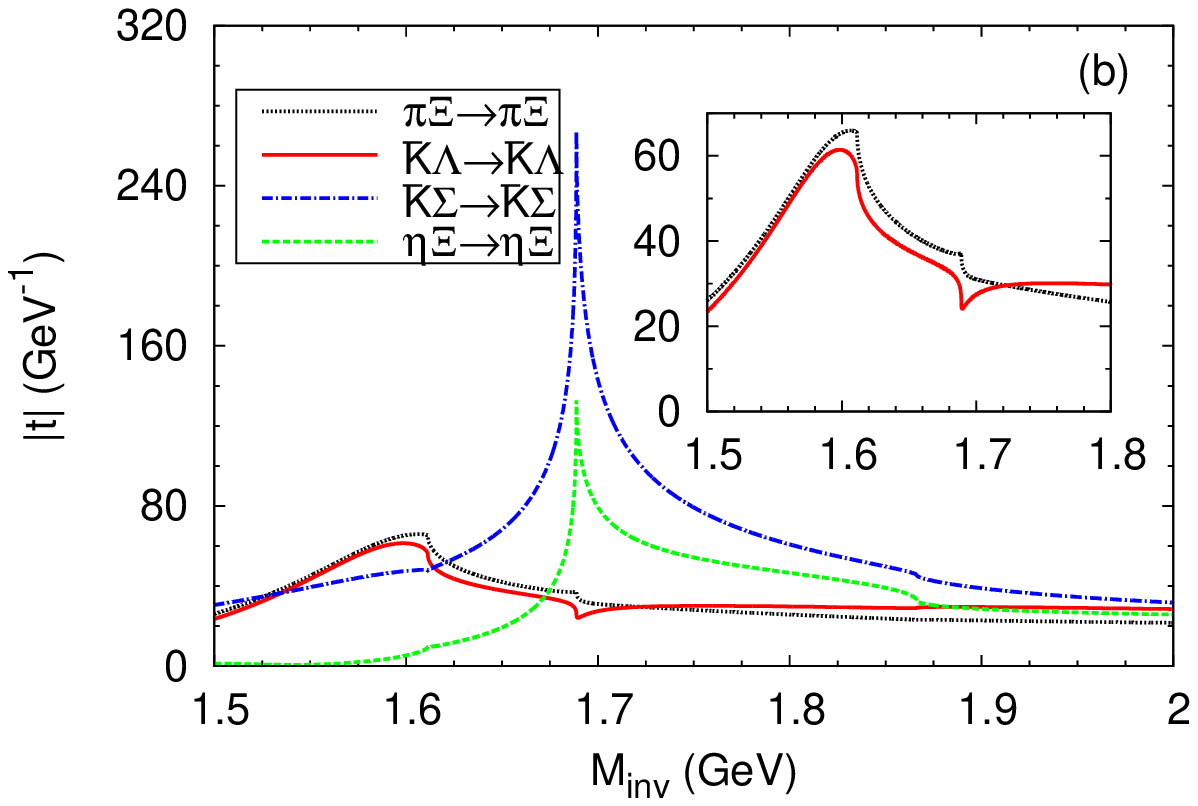}
\caption{(Color online) Absolute value of $t_{\pi\Xi,\pi\Xi}$, $t_{\bar{K}\Lambda, \bar{K}\Lambda}$, $t_{\bar{K}\Sigma, \bar{K}\Sigma}$, and $t_{\eta\Xi,\eta\Xi}$ as a function of the invariant mass of 
$\pi\Xi$, $\bar{K}\Lambda$ , $\bar{K}\Sigma$, $\eta\Xi$ system: (a) $a_l=-2$ for the channels of $\pi\Xi$, $\bar{K}\Lambda$, $\bar{K}\Sigma$, and $\eta\Xi$, and (b) the same as (a) but with 
$a_{\pi\Xi}=-3.1$ and $a_{\bar{K}\Lambda}=-1$.  The insets show the magnified $\pi\Xi\rightarrow\pi\Xi$ and $\bar{K}\Lambda\rightarrow\bar{K}\Lambda$ amplitudes. } 
  \label{fig:moduleT}
\end{figure}

Assuming the existence of the strangeness counterpart of the $P_c(4450)$ as shown in
Refs.~\cite{Wu:2010jy,Wu:2010vk} and following the approach
proposed in Ref.~\cite{Roca:2015dva} to describe the LHCb data, we can parameterize the transition  matrix element for 
$J/\psi\Lambda$ in Eq.~(\ref{eqn:fullamplitude})  as
\begin{equation}
t_{J/\psi\Lambda,J/\psi\Lambda}=\frac{g^2_{J/\psi\Lambda}}{M_{J/\psi\Lambda}-M_R+i\frac{\Gamma_R}{2}}.
\end{equation}
In Refs.~\cite{Wu:2010jy,Wu:2010vk}, two states are found in the strangeness $-1$ and isospin 0 channel with the following
pole positions $\sqrt{s}=4368-2.8i$ and $\sqrt{s}=4547-6.4i$. These numbers are obtained without any fine tuning. If now we assume that one of these states corresponds to
the strange counterpart of the $P_c(4450)$ and use the experimental measurement of its mass as a reference, we can imagine that the counterpart of the $P_c(4450)$, $P_{cs}$, should 
appear at $M_{P_c(4450)}+\Delta M$, where $\Delta M$ can be estimated using either the $\Lambda N$ or $\Sigma N$ mass difference, which are 175 and 257 MeV, respectively. 
As an rough estimate, one can take $\Delta M=200$ MeV and obtain $M_R=4650$ MeV. As for $\Gamma_R$, it should be of order 10 MeV~\cite{Wu:2010jy,Wu:2010vk}, and therefore we take $\Gamma_R=10$ MeV. For the coupling  $g_{J/\psi\Lambda}$, we use a value of $0.5\sim 0.6$,
as given in Ref.~\cite{Wu:2010vk}. 

Finally, the invariant mass distribution of the process $\Xi^-_b\to J/\psi\Lambda K^-$ reads
\begin{align}\label{eqn:dGammadM}
\frac{d\Gamma}{dM^2_{J/\psi\Lambda}dM^2_{\bar{K}\Lambda}}
=\frac{1}{(2\pi)^3}\frac{4M_{\Xi_b} M_\Lambda}{32M^3_{\Xi_b}} \left|\mathcal{M}(M_{J/\psi\Lambda}, M_{\bar{K}\Lambda})\right|^2\,,
\end{align}
where $M_{J/\psi\Lambda}$ and $M_{\bar{K}\Lambda}$ are the invariant masses of $J/\psi\Lambda$ and $\bar{K}\Lambda$. For a given value of $M_{J/\psi\Lambda}$, the range of $M^2_{\bar{K}\Lambda}$ is defined as,
\begin{eqnarray}
(M^2_{\bar K\Lambda})_{\rm max}\!&=&\! \left(E^*_\Lambda+E^*_{\bar K}\right)^2 \nonumber \\ && -\left(\sqrt{E^{*2}_\Lambda-M^2_\Lambda}-\sqrt{E^{*2}_{\bar{K}}-m^2_{\bar{K}}}\right)^2,  \nonumber \\ 
(M^2_{\bar K\Lambda})_{\rm min}\!&=&\! \left(E^*_\Lambda+E^*_{\bar K}\right)^2 \nonumber \\ &&-\left(\sqrt{E^{*2}_\Lambda-M^2_\Lambda}+\sqrt{E^{*2}_{\bar{K}}-m^2_{\bar{K}}}\right)^2,  \nonumber \\ 
\end{eqnarray} 
where $E^{*}_\Lambda=(M^2_{J/\psi\Lambda}-m^2_{J/\psi}+M^2_\Lambda)/2M_{J/\psi\Lambda}$ and $E^{*}_{\bar{K}}=(M^2_{\Xi_b}-M^2_{J/\psi\Lambda}-m^2_{\bar{K}})/2M_{J/\psi\Lambda}$ are the energies of $\Lambda$ and $\bar{K}$ in the $J/\psi\Lambda$ rest frame.
Similar formulas are obtained for the range of $M^2_{J/\psi\Lambda}$ when we fix $M_{\bar{K}\Lambda}$.

\begin{figure*}[!htb]
\centering
  \includegraphics[width=0.45\textwidth]{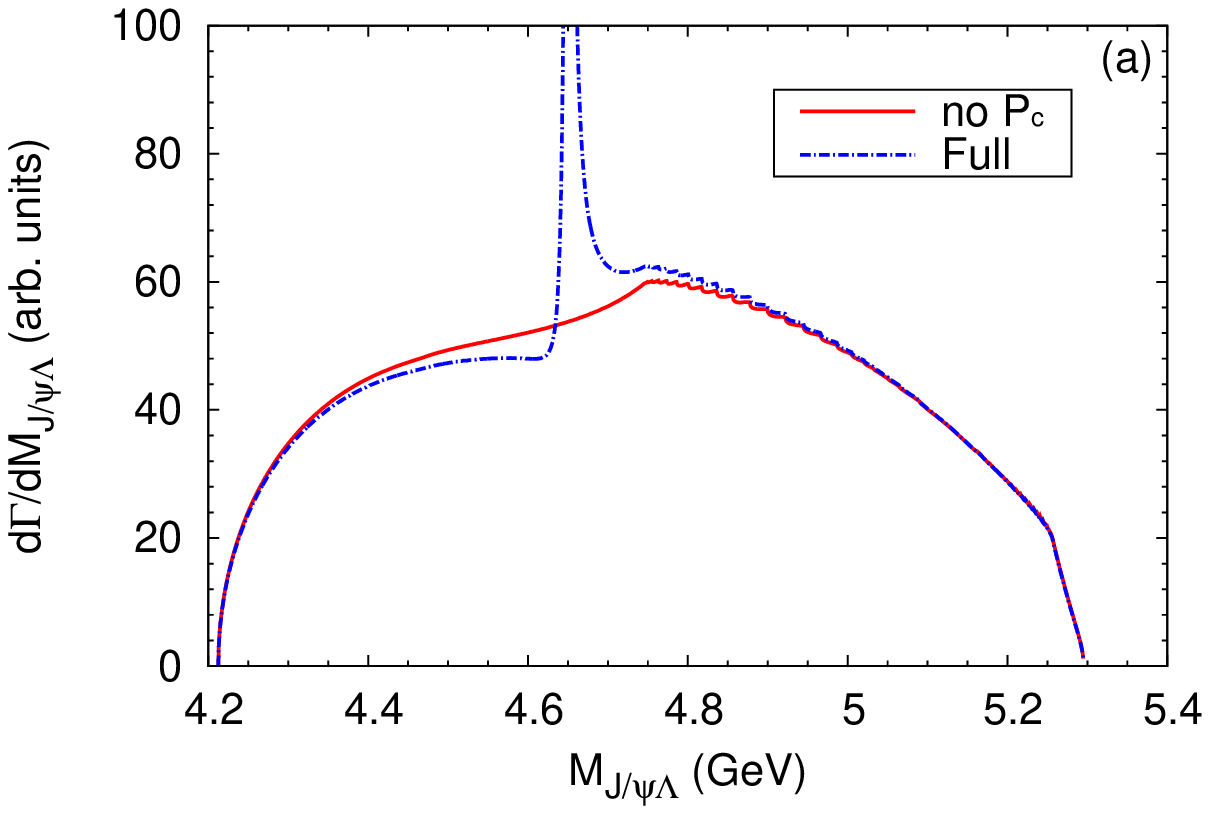}
    \includegraphics[width=0.45\textwidth]{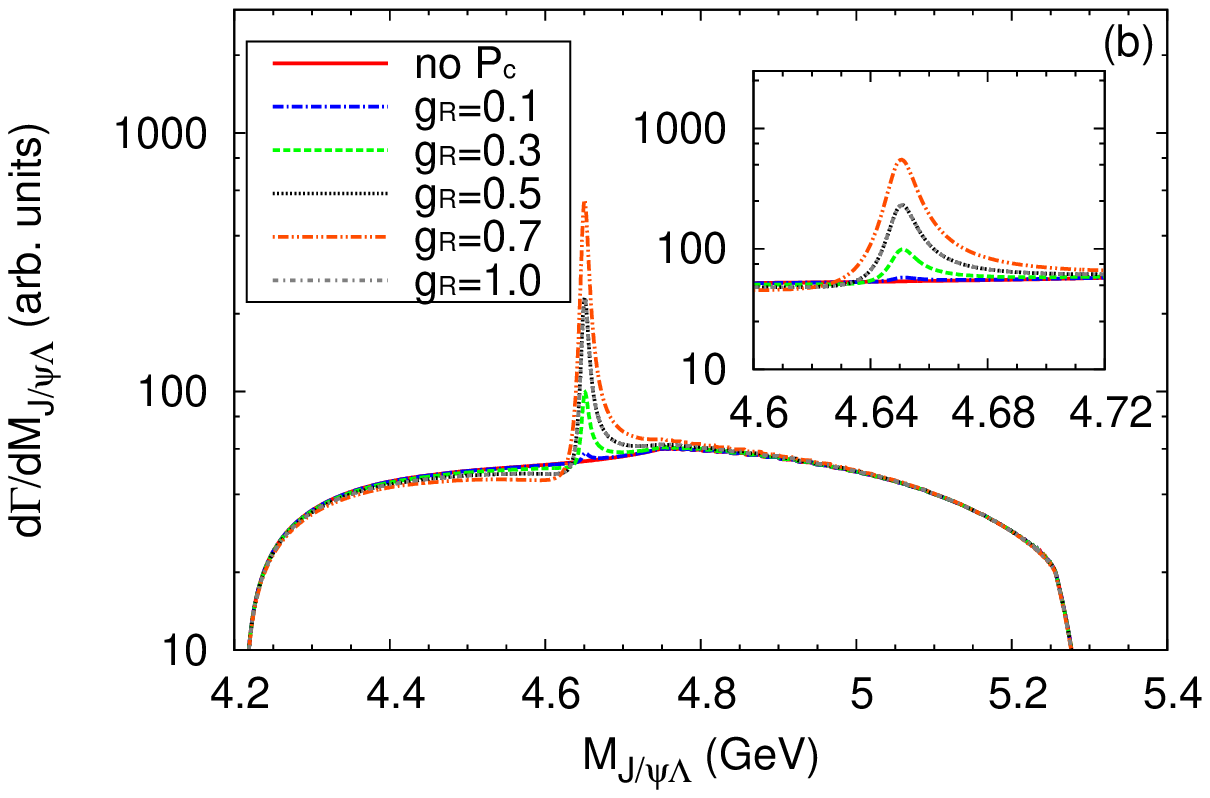}
      \includegraphics[width=0.45\textwidth]{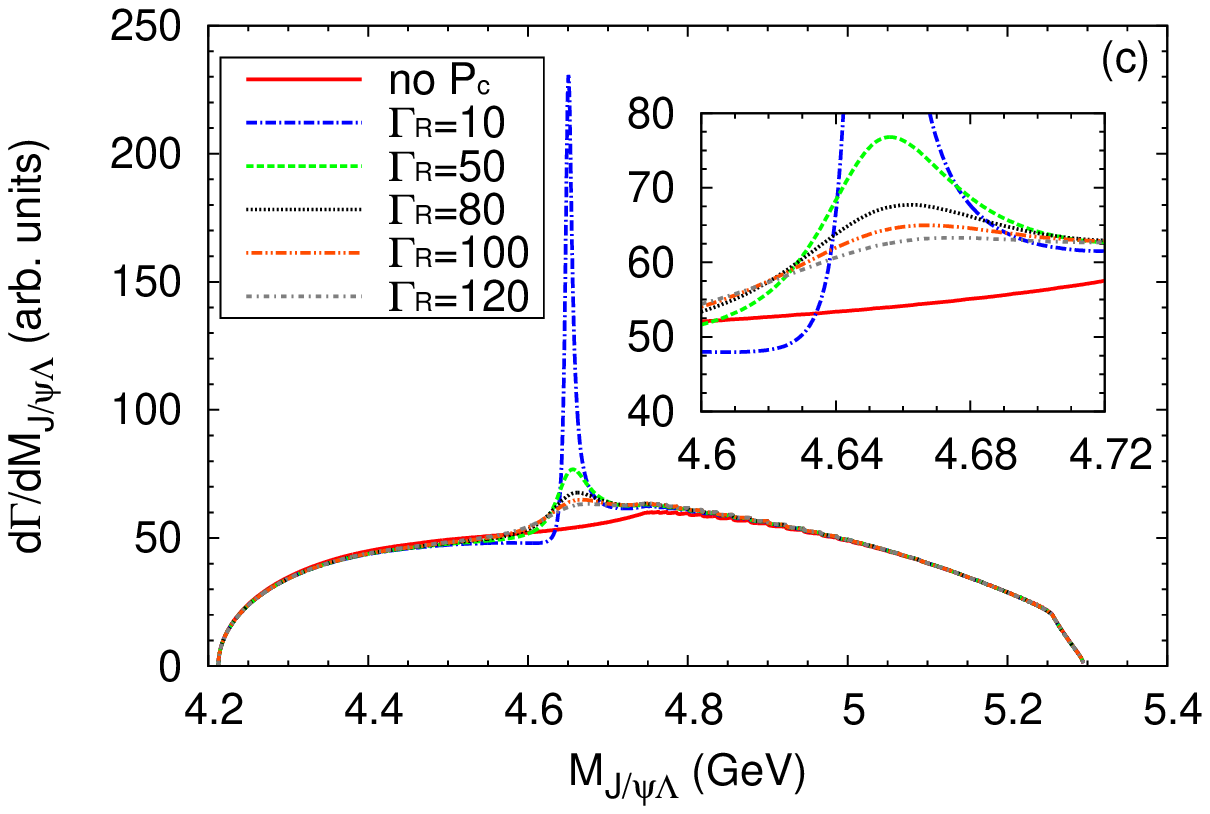}
        \includegraphics[width=0.45\textwidth]{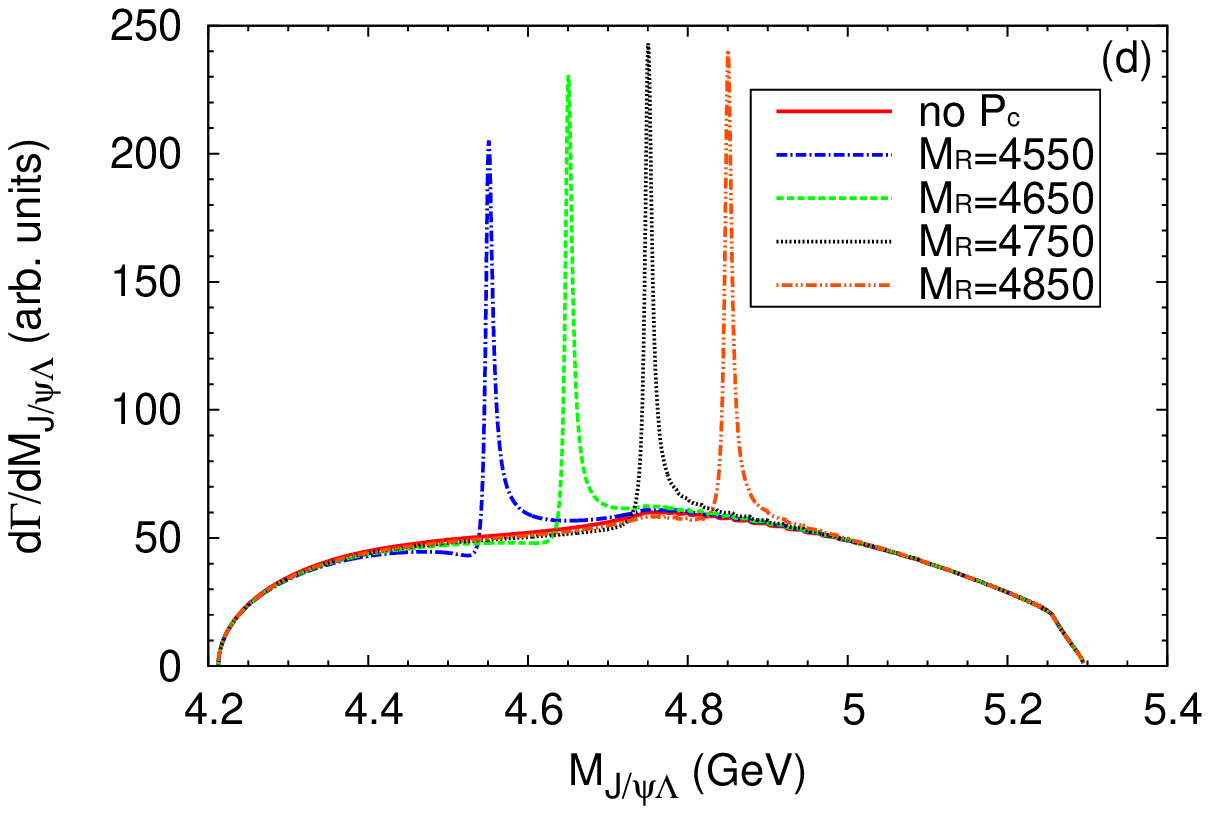}
  
\caption{(Color online) $J/\psi\Lambda$ invariant mass distributions for  $\Xi_b\to J/\psi \Lambda\bar{K}$ with $M_R=4650$ MeV, $g_{J/\psi\Lambda}=0.5$, and
$\Gamma_R=10$ Mev (a) and with two of the parameters fixed and the other varying as shown in the plot (b,c,d).} 
  \label{fig:dwidth}
\end{figure*}

\begin{figure}[!htb]
\centering
  \includegraphics[width=0.45\textwidth]{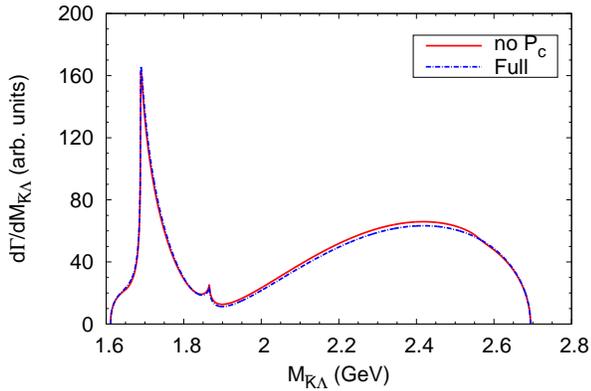}
\caption{(Color online) $\bar{K}\Lambda$ invariant mass distributions for $\Xi_b\to J/\psi \Lambda\bar{K}$ .} 
  \label{fig:dwidth2}
\end{figure}

\section{Results and Discussion}\label{sec:results}
In this section, we present our results for the process $\Xi^-_b\rightarrow J/\psi\Lambda K^-$. First, we show the absolute value of the transition amplitudes $|t|$ for the $\pi\Xi$, $\bar{K}\Lambda$, $\bar{K}\Sigma$ , and $\eta\Xi$ in $I=1/2$ and $S=-2$ in Fig.~\ref{fig:moduleT}(a). We also show the corresponding results with  $a_{\pi\Xi}=-3.1$ and $a_{\bar{K}\Lambda}=-1$ in Fig.~\ref{fig:moduleT}(b). The results for both choices look very similar.  
The $\Xi(1620)$ can be clearly seen in the $\pi\Xi$ invariant mass distributions, and $\bar{K}\Lambda$ distributions in Fig.~3(b), but not so prominently in Fig.~3(a). As a result,
we anticipate that an experimental study of the $\Xi^-_b\rightarrow J/\psi K^-\Lambda$ can yield valuable information on the poorly known $\Xi(1620)$ as well.

Next, we predict the $J/\psi\Lambda$  invariant mass distribution for the $\Xi_b\to J/\psi \Lambda\bar{K}$ decay in Fig.~\ref{fig:dwidth}, up to an arbitrary normalization ($V_p=1$).
The red solid line shows the result without the $J/\psi\Lambda$ interaction [only the term $h_{\bar{K}\Lambda}+T_{\bar{K}\Lambda}$ of Eq.~(\ref{eqn:fullamplitude})], and the blue dashed-dotted line stands for the result of our full model. We observe a prominent structure around 4650 MeV on top of the background when the $J/\psi\Lambda$ interaction is taken into account.  A variation of $M_R$ shifts the peak position accordingly, but
a clear signal can still be observed.  Furthermore, as long as the width is smaller than 100 MeV, experimental observation should not be too difficult.
 We stress however that the strength of the signal will depend strongly on the coupling of the hidden charm state to the $J/\psi \Lambda$, i.e., $g_{J/\psi\Lambda}$. A much smaller value of the coupling
will diminish the signal as naively expected.  Therefore, indeed an experimental study of the decay mode we propose can help to verify or disprove the unitary approach for this particular case.

In Fig.~\ref{fig:dwidth2}, we show the invariant mass distribution of the $\Xi^-_b$ decay as a function of the $\bar{K}\Lambda$ invariant mass. It is seen that the $J/\psi\Lambda$ final state interaction
does not affect much the predicted shape. However, the two cusps reflecting the $\bar{K}\Sigma$ and $\eta \Xi$ thresholds can be easily recognized.  
The strong enhancement around $M_{\bar{K}\Lambda}\approx1.7$ GeV  can be identified with the $\Xi(1690)$ as in Ref.~\cite{Gamermann:2011mq}. Depending on
the particular parameter set for the $\bar{K}\Lambda$ interaction, the $\Xi(1620)$ is also visible.

It is to be noted that the decay mechanism of the present process is the same as that of the $\Lambda_b^0\rightarrow J/\psi K^- p$~\cite{Roca:2015dva}. 
In both decays, the involved CKM matrix element is the product of $V_{cs}V_{cb}$. Therefore, as a crude estimate,
we would like to guess that the decay rate of the $\Xi^-_b\rightarrow J/\psi K^-\Lambda$ is of the same order of magnitude as that of $\Lambda_b^0\rightarrow J/\psi K^-\Lambda$, 
neglecting the difference in phase space and final state interactions.  This is somehow consistent with the study of Ref.~\cite{Cheng:2015cca}, where it is found that $\Gamma(\Xi^-_b\rightarrow P_\Lambda K^-)/\Gamma(
\Lambda^0_b\rightarrow P_p K^-)=0.408(0.343)$, where $P_{p/\Lambda}$ denotes a pentaquark state having the same light quark composition as that of the proton or $\Lambda$, while the first number
is for spin 3/2 and that in the parenthesis for spin 5/2. Therefore, both studies show that the decay mode proposed in the present work should and can be studied at LHCb.

\section{Summary}

We have proposed to study the $\Xi^-_b\rightarrow J/\psi  K^- \Lambda$ decay to measure  a hidden-charm pentaquark state with strangeness, predicted to
exist in the unitary approach. This model has  predicted the existence of two non-strange hidden-charm pentaquark states in the energy region where the
$P_c(4450)$ has been seen. The decay mechanism we employed has been previously adopted to successfully describe the LHCb $\Lambda_b^0\rightarrow J/\psi K^- p$ invariant mass distributions. 
Our study showed that the strange hidden-charm pentaquark state can be clearly seen on top of the background. Given the fact that both the unitary model and the reaction mechanism have been
tested in describing the LHCb $\Lambda^0_b$ decay, we strongly encourage our experimental colleagues to study the $\Xi^-_b\rightarrow J/\psi K^- \Lambda$ decay  proposed here, which can be very helpful to test the existence of the
pentaquark states and their nature.

\section{Acknowledgements}
One of us, E. O., wishes to acknowledge support from the Chinese
Academy of Science (CAS) in the Program of Visiting Professorship
for Senior International Scientists (Grant No. 2013T2J0012). L.S.G thanks the Institute for Nuclear Theory at University of Washington
for its hospitality and the Department of Energy for partial support during the completion of this work. This work is partly supported by the
National Natural Science Foundation of China under Grant Nos.
11475227, 1375024, 11522539, 11505158, 11475015, and 11165005, the Open Project Program of State
Key Laboratory of Theoretical Physics, Institute of Theoretical
Physics, Chinese Academy of Sciences, China (No.Y5KF151CJ1), the Spanish Ministerio de Economia y
Competitividad and European FEDER funds under the contract number
FIS2011-28853-C02-01 and FIS2011-28853-C02-02, and the Generalitat
Valenciana in the program Prometeo II-2014/068. We acknowledge the
support of the European Community-Research Infrastructure
Integrating Activity Study of Strongly Interacting Matter (acronym
HadronPhysics3, Grant Agreement n. 283286) under the Seventh
Framework Programme of EU.

\end{document}